\documentclass[doublecol]{epl2_arXiv} 

\usepackage{graphicx}
\usepackage{amsmath}
\usepackage{amssymb}
\usepackage{amsfonts}
\usepackage{mathtools}
\usepackage{bm}
\usepackage{nicefrac}
\usepackage{hyperref}
\usepackage[caption=false,position=top]{subfig}
\usepackage[english]{babel}

\hypersetup{
	colorlinks   = true,     
	urlcolor     = black, 
	linkcolor    = blue, 
	citecolor  	 = blue  
}

\def\InGaAs{In$_x$Ga$_{1-x}$As~}
\def\Bext{B_\mathrm{ext}}
\def\Bov{\B_\mathrm{Ov}}
\def\Beff{B_\mathrm{eff}}
\def\Omegaeff{\Omega_\mathrm{eff}}
\def\ddt{\frac{\mathrm{d}}{\mathrm{d}t}}
\def\ex{\bm{e}_x}
\def\ez{\bm{e}_z}
\def\S{\bm{S}}

\def\B{\bm{B}}
\def\TR{T_\mathrm{R}}
\def\Tnstar{T_\mathrm{n}^\ast}
\def\T2star{T_\mathrm{2}^\ast}
\def\np{n_\mathrm{p}}
\def\ge{g_\mathrm{e}}
\def\gh{g_\mathrm{h}}
\def\muB{\mu_\mathrm{B}}
\def\SSML{S_\mathrm{SML}}

\title{Nuclear magnetic resonance spectroscopy of nonequilibrium steady states in quantum dots}
\shorttitle{NMR spectroscopy of nonequilibrium steady states in quantum dots}

\author{P.~Schering\thanks{E-mail: \email{philipp.schering@tu-dortmund.de}} \and G.~S.~Uhrig\thanks{E-mail: \email{goetz.uhrig@tu-dortmund.de}}}
\shortauthor{P. Schering \etal}

\institute{
	Condensed Matter Theory, TU Dortmund University - D-44221 Dortmund, Germany
}


\abstract{
The optically induced polarization of localized electron spins in an ensemble of quantum dots~(QDs) dephases due to the interaction with the surrounding nuclear spins. Despite this dephasing, the spins in the QDs can be controlled to respond coherently by applying periodic laser pulses, leading to a revival of the spin polarization before each pulse. This effect, known as spin mode locking, strongly depends on an emerging selection of certain polarizations of the nuclear spin bath which is driven to a steady state far from equilibrium. We investigate the influence of the nuclear composition in In$_x$Ga$_{1-x}$As~QDs on this nonequilibrium behavior and demonstrate that nuclear magnetic resonances~(NMR) appear as very sharp minima in the magnetic field dependence of the revival amplitude. This suggests a novel kind of NMR spectroscopy.
}

\begin{document}

\maketitle

\section{Introduction}
The coherent manipulation of localized electron spins in quantum dots~(QDs) and their nuclear environment is 
a highly active and promising route to quantum information~\cite{divincenzo98}. Only recently, it was demonstrated that the nuclear environment can be employed as persistent quantum memory~\cite{gangloff19,denning19}.
Fascinatingly, the hyperfine interaction between the electron spins and the nuclei in the QDs can be employed in several other ways, e.\,g., by driving the $10^4 - 10^6$ nuclei in a single QD~\cite{urba13} to act coherently through the application of long trains of periodic laser pulses. By means of spin mode locking~(SML)~\cite{greil06b}, where 
 the spin polarization revives before each next pulse, it is possible to overcome dephasing constraints in an ensemble of QDs. 
This effect is strongly influenced by nuclei-induced frequency focusing~(NIFF)~\cite{greil07a}. The electron spin sees an effective field being the superposition of the applied transverse magnetic field and the field exerted by the nuclear spins. Periodic application of laser pulses drives the system to a nonequilibrium steady state~(NESS) in which the precession period of the electron spin is commensurate with the pulse repetition period.
Through tailored pulse protocols it is in fact possible to generate a coherent single-mode spin precession of the full QD \emph{ensemble}~\cite{greil09,evers20} because NIFF, instead of the ordinary dynamic nuclear polarization,
strongly prolongates the coherence time. This allows for the implementation of ulrafast optical rotations of electron spins localized in QDs~\cite{greil09b}, which is a key prerequisite for quantum computation.
Many experiments have been conducted on this 
issue~\cite{glazov10,spatzek11,varwig14,jasch17,klein18,evers18,markmann19} revealing its complexity and the associated difficulty of a thorough theoretical description.

In these experiments, electron spin polarization is periodically induced in the QD ensemble along the axis of the laser beam ($z$~axis) by the optical excitation of trion states (transition energy $\approx 1.4\,$eV) while a strong transverse magnetic field is applied (Voigt geometry)~\cite{greil06a}; see ref.~\cite{glazo12b} for a review. Eventually, the spin polarization of the electrons is transferred to the nuclei via the hyperfine interaction.
The coherence time of the localized electron spin is of the order of nanoseconds while the nuclei have a much longer coherence time up to minutes~\cite{greil07a}. Thus, the NESS imprinted into the nuclear spin bath lives on a macroscopic timescale.

The goal of the present Letter is to describe QDs which comprise several different isotopes and hence
different species of nuclear spins. Unexpectedly, we find very sharp signatures in the revival amplitudes
which result from nuclear magnetic resonances, more precisely from the commensurability of the nuclear Larmor period and the repetition time of the applied pulses. This suggests a novel kind of nuclear magnetic resonance~(NMR) spectroscopy. We stress that this study crucially relies on theoretical progress in the implementation of the simulations since the demanding run time resources cannot be provided otherwise.

Recent theoretical developments have established a sophisticated semiclassical model to describe the interplay of SML and NIFF~\cite{scher20} by combining ideas from previous research~\cite{yugov09,yugov12,glazo12a,jasch17,scher18,klein18,scher19}. The qualitative physics is already captured in a slightly less complex model~\cite{scher20}. In this Letter, we extend this model to account for several species of nuclei in \InGaAs QDs.
Up to now, the role of several nuclear species has only been studied in a perturbative quantum mechanical model for a small number of nuclear spins~\cite{beuge17}. This does not allow to study the NESS present in experiments.
In other previous studies~\cite{klein18,scher20}, only a single average nuclear spin is considered, reducing the complexity by a significant amount.
The present extended model enables us to study the influence of the QD composition on the 
nonequilibrium behavior. We find a remarkable sensitivity to nuclear magnetic resonances.

\section{Model}
The spin dynamics in a single QD is governed by the hyperfine interaction between the localized electron spin and the surrounding nuclei in conjunction with an transverse magnetic field inducing Larmor 
precession~\cite{khaetskii02,merku02,hanson07,urba13}.
We apply the recently developed minimal semiclassical model from ref.~\cite{scher20} and extend it to several nuclear species in \InGaAs QDs. 
Besides the internal dynamics, the quantum mechanical action of laser pulses leading to spin orientation needs to be described.

\subsection{Spin dynamics}
The spin dynamics is described by the precession equations ($\hbar$ set to unity)
\begin{subequations}
	\begin{align}
	\ddt \S &= ( \Bov + \gamma_\mathrm{e} B_\mathrm{ext} \ex ) \times \S + \frac{1}{\tau_0} J^z \ez \,,\\
	\ddt \bm{B}_{\mathrm{Ov},k} &= ( A_k \S + \gamma_{\mathrm{n},k} B_\mathrm{ext} \ex) \times \bm{B}_{\mathrm{Ov},k} \,.
	\end{align}
	\label{eq:EoM_isotopes}%
\end{subequations}
The electron spin $\S$ precesses in the superposition of the time-dependent Overhauser field $\Bov = \sum_{j=1}^{N} A_j \bm{I}_j$,
which is the sum of all nuclear spins weighted by their hyperfine coupling, and the transverse external magnetic field $\gamma_\mathrm{e} \Bext \ex$, where $\gamma_\mathrm{e} = \ge \muB$ denotes the gyromagnetic ratio, $\ge = 0.555$~\cite{greil07a} the electronic $g$ factor, and $\muB$ the Bohr magneton.
The total Overhauser field is the sum of the subfields of each nuclear species, i.\,e., 
$\Bov = \sum_k \bm{B}_{\mathrm{Ov},k}$.
Each subfield precesses in the Knight field $A_k \S$ due to the hyperfine interaction with the electron spin and in the external field. The gyromagnetic ratios relevant for the influence of the external field are very small, 
$\gamma_{\mathrm{n},k}/\gamma_\mathrm{e} = \mathcal{O}(10^{-3})$~\cite{coish09}. 
Yet, since the energy scale $\gamma_{\mathrm{n},k}\Bext$ can be of similar order as $A_k$, the nuclear Zeeman effect turns out to determine the nonequilibrium spin physics decisively.
The strength of the hyperfine interaction is proportional to the probability density of the electron at the position of the nuclei. For simplicity, we approximate this density as a box with a certain spatial cutoff, i.\,e., we use
the so called box model~\cite{merku02}. Then, all nuclear spins of species $k$ share the same coupling constant~$A_k$.
This is the key difference to ref.~\cite{scher20} where a more realistic exponential parametrization of the hyperfine couplings is considered for a single nuclear spin species. Using the box model, however, does not change the qualitative interplay of SML and NIFF~\cite{scher20}.

The optically excited trion state consists of two electrons in a spin singlet and a heavy hole with unpaired spin, which can be captured by a pseudospin~\cite{glazo12b}. The dynamics of its $z$ projection is given by
\begin{align}
J^z (t) = J^z(0)\, \mathrm{e}^{-t/\tau_0} \cos(\gamma_\mathrm{h} \Bext t) \,,
\end{align}
where $\gh = \gamma_\mathrm{h}/\muB = 0.15$~\cite{yugov07} is the $g$ factor of the heavy hole in the trion.
The pseudospin precesses in the transverse magnetic field while the trion decays radiatively into the ground state, characterized by the electron spin $\S$. The radiative trion lifetime is 
\mbox{$\tau_0 = 400\,$ps}~\cite{greil06a,greil06b}. 
The hyperfine interaction is a magnitude smaller for heavy holes than for electrons and also strongly 
anisotropic~\cite{fischer08,testelin09,zhukov18}. 
Since this was found to barely affect the physics~\cite{scher20}, we neglect it here for simplicity.

Due to the large number of nuclei in a QD ($10^4-10^6$), the Overhauser field can safely be treated as a classical field based on the central limit theorem~\cite{merku02,stanek14}. 
Using the truncated Wigner approximation~\cite{polkovnikov10}, the classical equations of motion~(EoMs)~\eqref{eq:EoM_isotopes} are solved for $M$ random initial configurations of the Overhauser field and of the electron spin
and they are averaged over the classical trajectories to approximate the quantum mechanical time evolution of the electron spin $\S$.
For the temperature of about $6$\,K prevalent under experimental conditions~\cite{greil06a,greil07a}, the nuclear spin bath is completely disordered and its fluctuations follow a normal distribution.
To capture its variance, we define the dephasing time 
\begin{align}
\Tnstar \coloneqq \sqrt{2} \left( \sum_{j=1}^N \frac{I_j(I_j + 1)}{3}A_j^2\right)^{-1/2} \,,	
\label{eq:Tnstar_general}
\end{align}
based on the hyperfine interaction of the electron with $N$ nuclei with spin $I_j$.
We define the ratios $\alpha_k$ by $A_k = \alpha_k A$ relative to $A \coloneqq A_1$.
Moreover, $N_k = n_k N$ is the number of nuclei of species $k$ so that $n_k$ describes 
its relative abundance in the \InGaAs QD. Then, we have
\begin{align}
A = \frac{\sqrt{2}}{\Tnstar \sqrt{N}} \left( \sum_k n_k \alpha_k^2 \frac{I_k(I_k+1)}{3} \right)^{-1/2} \,, \label{eq:energyscale_A}
\end{align}
which is the natural energy scale here, depending on $\Tnstar$, $N$, sample specific parameters, and 
known properties of the various nuclear species~\cite{coish09}.
Using the quantum mechanical second moment of a disordered nuclear spin, 
the variances of the nuclear subfields read
\begin{align}
\mathrm{Var}\left[ B^\alpha_{\mathrm{Ov},k} \right] = 
\sum_{j=1}^{N_k} \frac{I_j(I_j + 1)}{3} A_k^2 = n_k N \alpha_k^2 A^2 \frac{I_k(I_k+1)}{3} \,,
\end{align}
$\alpha \in \{x,y,z\}$.
Since variances are additive, the fluctuations of the Overhauser field can be characterized by the dephasing time $\Tnstar$ via
\begin{align}
\mathrm{Var}\left[ B^\alpha_{\mathrm{Ov}} \right] = \sum_{j=1}^{N} \frac{I_j(I_j + 1)}{3} A_j^2 
= \frac{2}{(\Tnstar)^2} \,. 
\label{eq:Overhauser_variance}
\end{align}
Thus, $\Tnstar$ is an input in our simulations taken from experiments~\cite{urba13}. 
We focus on a QD sample with \mbox{$\Tnstar = 4\,$ns}~\cite{greil06a,fischer18}.
This dephasing time captures only the dephasing due to the fluctuating nuclear spins. 
In an inhomogeneous ensemble of QDs, the actually observed dephasing time $\T2star$ shows a strong magnetic field dependence due to a spread of the electronic $g$ factors~\cite{greil06a,fischer18}. 
We do not include this inhomogeneity because it turned out to be irrelevant for the NESS~\cite{scher20}.
In the simulations, we use $N = 60$ nuclear spins since the number of pulses and therefore also the run time required to 
reach the NESS scales linearly with $N$ because larger $N$ implies smaller $A_k$. Note that the
number of equations does not depend on $N$ in the applied box model because all nuclear spins of the same species precess with identical frequencies.
We checked, however, that a larger $N$ only has small influence 
on the results, see Supplementary Material.
The parameters of the nuclei are listed in table~\ref{tab:nuclei_parameters}.
We consider an average isotope for $^{113}$In and $^{115}$In weighted by their natural abundances
and denoted by $\overline{\mathrm{In}}$ because their parameters are almost identical.
Eventually, we consider four different nuclear spin species.

\begin{table}[h!]
	\centering
	\caption{Nuclear spin $I$, gyromagnetic ratio $\gamma_\mathrm{n}$, and hyperfine interaction strength $A_\mathrm{hf}$ for the stable isotopes in \InGaAs QDs according to ref.~\cite{coish09}. Only the ratios of the hyperfine couplings enter in our model.}
	\label{tab:nuclei_parameters}
	\begin{center}
		\begin{tabular}{l|rrrr}
			isotope & $I$ & $\gamma_\mathrm{n}$ (rad\,T$^{-1}$\,s$^{-1}$) & $A_\mathrm{hf}$ ($\mu$eV)\\
			\hline
			$^{69}$Ga	&	$3/2$	& $6.43 \cdot 10^7$	& 74	\\
			$^{71}$Ga	&	$3/2$	& $8.18 \cdot 10^7$	& 96	\\
			$^{75}$As	&	$3/2$	& $4.60 \cdot 10^7$	& 86 	\\
			$^{113}$In	&	$9/2$	& $5.88 \cdot 10^7$	& 110	\\
			$^{115}$In	&	$9/2$	& $5.90 \cdot 10^7$	& 110	
		\end{tabular} 
	\end{center}
\end{table}

\subsection{Spin polarization induced by a laser pulse}
The QDs are exposed to trains of millions of circularly polarized pump pulses with repetition time ${\TR = 13.2}\,$ns~\cite{greil07a}, periodically exciting trion states~\cite{yugov09}. The radiative trion decay is restricted by the optical selection rules and in this way, the application of the pulses in a strong transverse field polarizes 
the electron spin~\cite{glazo12b}.
The typical pulse duration is about $2\,$ps~\cite{greil07a,klein18,evers18} which is an order of magnitude smaller than the Larmor period of the electron spin in a magnetic field of $10\,$T. Thus, the pump pulse can be taken to be instantaneous. We consider each pump pulse as a quantum mechanical measurement and account for the 
uncertainty principle~\cite{scher18,scher20}.
In the framework of a truncated Wigner approximation~\cite{polkovnikov10}, this implies that a nondeterministic pulse description is required to take first order quantum corrections into account.
We consider the application of resonant $\pi$~pulses with $\sigma^-$~helicity.
Then, the spin components before~($\S_\mathrm{b}$) and after~($\S_\mathrm{a}$) the pulse are normally distributed with expectation values and variances fulfilling~\cite{yugov09,scher20}
\begin{subequations}
	\begin{align}
	\mathrm{E}[S^z_\mathrm{a}] &= \frac{1}{4} + \frac{1}{2} S^z_\mathrm{b} \,, \\
	\mathrm{E}[S^x_\mathrm{a}] &= \mathrm{E}[S^y_\mathrm{a}] = 0 \,, \\
	\mathrm{Var}[S^\alpha_\mathrm{a}] &= \begin{cases} \frac{1}{4} - \mathrm{E}^2[S_\mathrm{a}^\alpha]\,, \qquad &\tx{if } \mathrm{E}^2[S^\alpha_\mathrm{a}] \le \frac{1}{4} \,, \\ 0\,, \qquad &\tx{else\,.} \end{cases}
	\end{align}
	\label{eq:pulse_distribution}%
\end{subequations}
Note that the trion has completely decayed before the next pump pulse ($\tau_0 \ll \TR$) and thus, 
the relation ${J^z_\mathrm{a} = S^z_\mathrm{b} - S^z_\mathrm{a}}$ holds due to spin conservation~\cite{yugov09}.

\subsection{Simulation details}
The numerical integration of the EoMs~\eqref{eq:EoM_isotopes} seems to be straightforward, but it requires massive parallelization to solve them $M = \mathcal{O}(10^4)$ times; we use $M=32512$. Moreover, reaching the experimental steady states for realistic parameters needs hundred millions of pulses, rendering a direct simulation elusive.
On the one hand, the fast Larmor frequency of the electron spin must be resolved numerically, so
only small time steps are possible. On the other hand, the time to reach the NESS scales 
$\propto\Bext^2$~\cite{scher18,scher20}. Eventually, the computational effort scales 
worse than $\propto \Bext^3$.
For this reason, we split the fast and the slow spin dynamics in the EoMs~\eqref{eq:EoM_isotopes}, 
solve the fast part analytically and expand the slow dynamics in first order of 
$\Bext^{-1}$, see Supplementary Material. This expansion is remarkably accurate for magnetic fields 
$\Bext \gtrsim 1\,$T while the numerical integration of the slow dynamics is two orders of magnitude faster
than the brute force approach. This methodical achievement is the key element which allows us to study 
the influence of several nuclear species on the nonequilibrium spin physics in QDs.

\section{Nonequilibrium spin physics}

\begin{figure}[t!]
	\centering
	\includegraphics[width=\columnwidth]{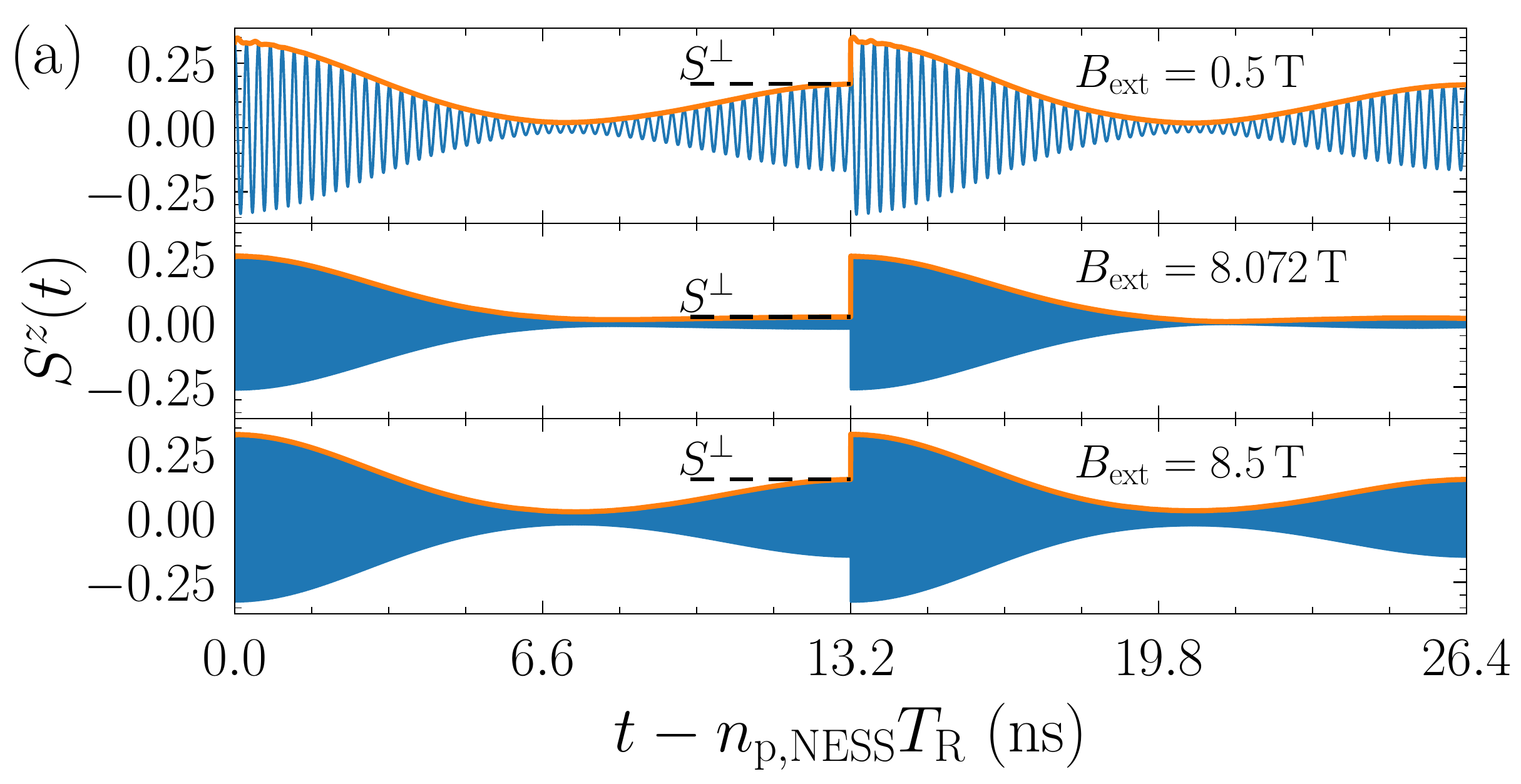}\\
	\includegraphics[width=\columnwidth]{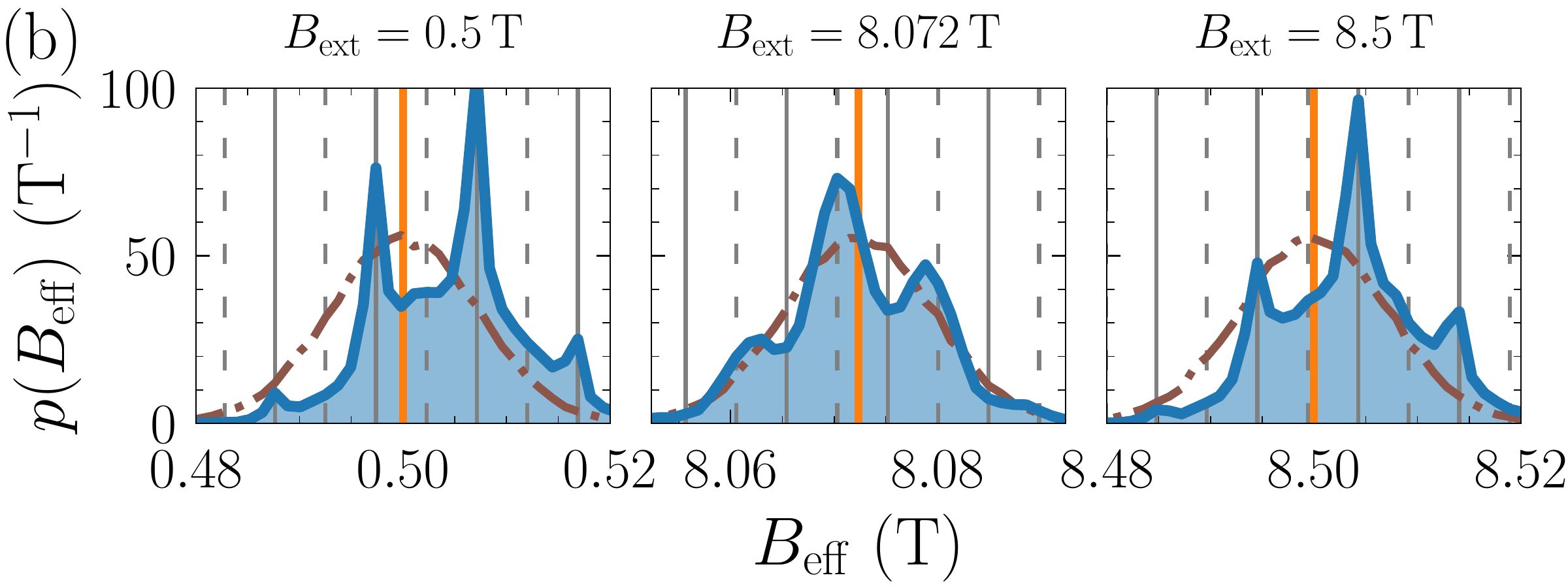}
	\caption{(a)~Electron spin dynamics showing spin mode locking in the NESS after $n_\mathrm{p,NESS} = 4.5 \cdot 10^5$~($\Bext = 	0.5\,$T), $1.174 \cdot 10^8$~($\Bext = 8.072\,$T), and $1.301 \cdot 10^8$~($\Bext = 8.5\,$T) pulses. The envelope is shown in orange, the fast Larmor precession is not discernible for large fields. 
		A pump pulse arrives at $0$ and $13.2\,$ns, respectively. The In concentration is $x=0.3$.
		(b)~Corresponding nonequilibrium probability distribution $p(\Beff)$ of the effective magnetic field (external plus Overhauser field) revealing the selection of special polarizations in the nuclear spin bath. The solid and dashed gray lines represent the phase synchronization conditions~\eqref{eq:ERC} and~\eqref{eq:ORC} for the electron spin, respectively. The orange vertical line highlights the external field $\Bext$. The initial normal distribution is depicted in brown for comparison.} 
	\label{fig:SpinDynamics_Overhauser}
\end{figure}

The nonequilibrium spin dynamics in QDs can be studied experimentally by time-resolved pump-probe spectroscopy where the Faraday rotation or ellipticity is measured yielding a signal \mbox{$\propto(S^z - J^z)$}~\cite{greil06a,yugov09}.
Since the trion decays quickly within $\tau_0 = 400\,$ps, $J^z$ does not contribute to the signal probed immediately before each next pump pulse.
After a long train of pump pulses, the spin dynamics $S^z(t)$ shows the typical behavior displayed in 
fig.~\ref{fig:SpinDynamics_Overhauser}(a).
The optically induced spin polarization is precessing around the external field $\Bext$ and 
dephases within~$\Tnstar = 4\,$ns.
Already $\mathcal{O}(10)$ pulses induce a revival of the spin polarization with amplitude $S^\perp = \nicefrac{1}{\sqrt{3}} - \nicefrac{1}{2} \mathrel{=\!\!\mathop:} \SSML$~\cite{beuge16,jasch17,klein18,scher20} before the next pulse as a consequence of SML in the random Overhauser field~\cite{greil06b,yugov12}.
The figure shows that applying many more pulses (for seconds to minutes in experiments) changes the amplitude $S^\perp$ of the revival due to NIFF~\cite{greil07a}; eventually a NESS is approached.
Generically, the amplitude is enhanced by the NIFF, but for certain conditions a destructive effect occurs~\cite{scher20}. 
For instance, for $\Bext = 8.072\,$T there is barely any revival, but for the slightly larger field 
$\Bext = 8.5\,$T the revival has a substantial amplitude.

The origin of the NIFF, partly responsible for the revival amplitude, 
is visualized in fig.~\ref{fig:SpinDynamics_Overhauser}(b)
by the probability distribution $p(\Beff)$ of the effective magnetic field 
\begin{align}
	\Beff \coloneqq | \Bov + \gamma_\mathrm{e} \Bext \ex | / \gamma_\mathrm{e} \, ,
\end{align}
Initially, due to the central limit theorem, it is a distribution centered around $\Bext$ 
with the variance given by eq.~\eqref{eq:Overhauser_variance}. 
But the long sequences of pulses train the system by preferring
commensurate modes so that the distribution evolves towards a comb-like structure until it reaches a NESS.
Hence, certain polarizations favoring commensurate dynamics are selected.
The position of the maxima in the NESS fulfill the even or the odd phase synchronization conditions 
for the Larmor precession of the electron spin~\cite{jasch17,klein18,scher20},
\begin{subequations}
	\begin{align}
	\Omegaeff \TR &= 2 \pi n \,, \label{eq:ERC} \\
	\Omegaeff \TR &= (2n + 1) \pi \,, \label{eq:ORC} 
	\end{align} \label{eq:resonances}%
\end{subequations}
$n \in \mathbb{Z}$, where $\Omegaeff = \gamma_\mathrm{e} \Beff$ is the precession frequency 
in the effective magnetic field $\Beff$, being the sum of the external field and the Overhauser field.
Condition \eqref{eq:ERC} is called `even' while condition \eqref{eq:ORC} is called `odd'.
The precise behavior of NIFF and its impact on the revival amplitude strongly depends on the applied magnetic field. Typically, the maxima represent even resonances~\eqref{eq:ERC}. 
But it was found that the system tends towards the odd resonance~\eqref{eq:ORC} instead
if the nuclear resonance condition
\begin{align}
	\gamma_{\mathrm{n},k} \Bext \TR = 2 \pi n \,,
	\label{eq:NRC_even}
\end{align}
$n \in \mathbb{Z}$, is met~\cite{scher20}. This is the case for $\Bext = 8.072\,$T shown in 
fig.~\ref{fig:SpinDynamics_Overhauser}, which is the resonant field for the nuclear spin of In for $n=1$. 
Then, the Larmor period of the In nuclear spins equals the pulse repetition time $\TR$ and the revival amplitude 
shows a sharp local minimum as function of the external field.

If a nuclear spin revolves an odd multiple of half Larmor precessions between two consecutive pulses, i.\,e.,
if
\begin{align}
\gamma_{\mathrm{n},k} \Bext \TR = (2n + 1) \pi \,, \label{eq:NRC_odd}
\end{align}
$n \in \mathbb{Z}$, holds, the revival amplitude as a function of the external field shows a 
very broad local minimum, see fig.~\ref{fig:Sperp_Bext_np}b around $\Bext = 4.5\,$T.
Here, the effective field still preferably realizes the even resonance~\eqref{eq:ERC}~\cite{scher20}.
So far, such an analysis has been conducted only for a single, average nuclear spin~\cite{scher20}. 
Below, we analyze the role of simultaneously present different nuclear species~$k$, 
giving rise to several possible resonance conditions and thus, increasing the complexity substantially.

\section{Nuclear magnetic resonance spectroscopy}
\begin{figure}[t!]
	\centering
	\includegraphics[width=\columnwidth]{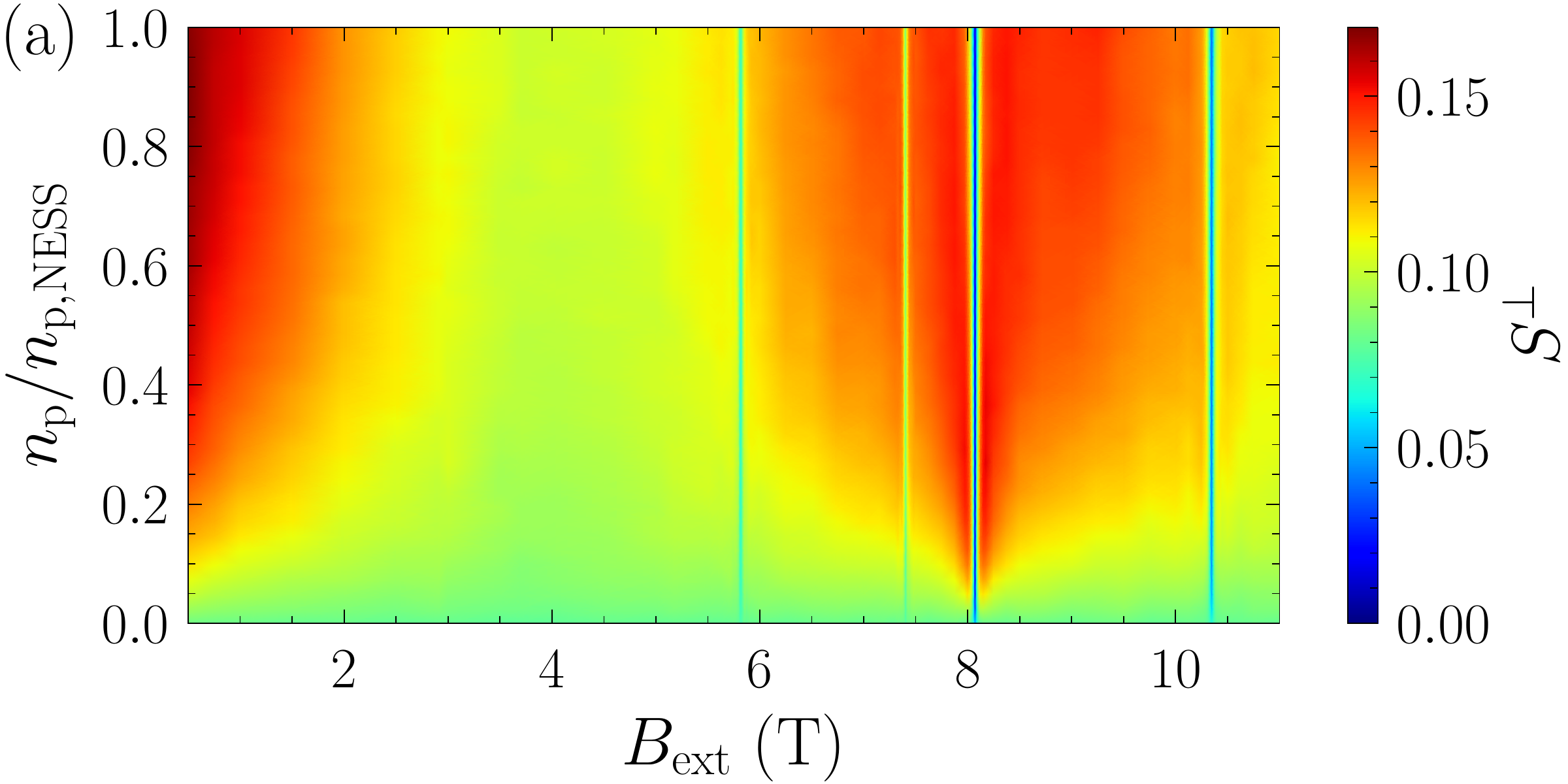}\\
	\includegraphics[width=\columnwidth]{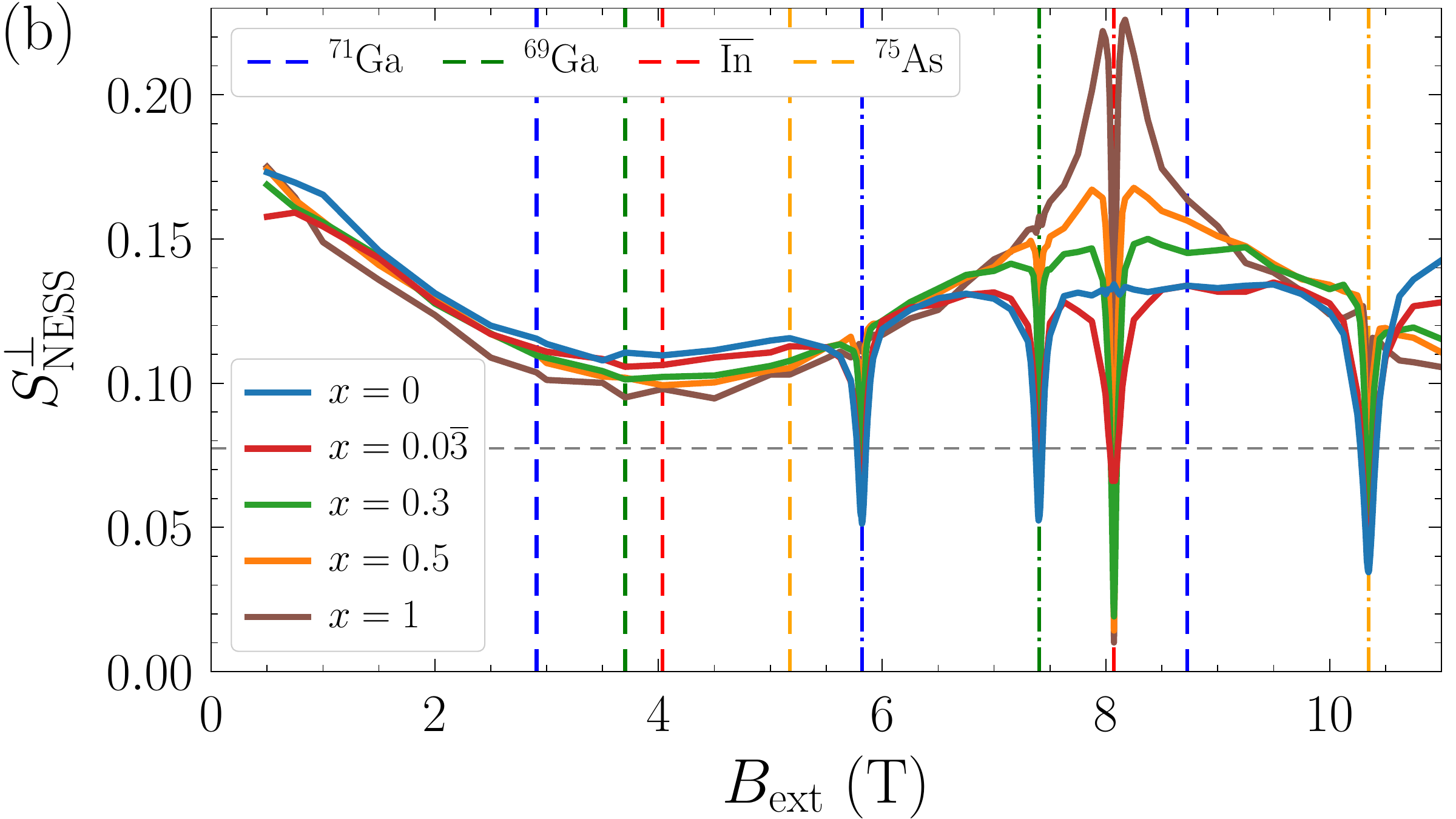}
	\caption{(a)~Revival amplitude $S^\perp$ as function of the external magnetic field $\Bext$ and of 
		the normalized number of pulses $\nicefrac{\np}{n_\mathrm{p,NESS}}$, where $n_\mathrm{p,NESS} \propto \Bext^2$ denotes the number of pulses necessary to reach the NESS. The In concentration is $x=0.3$. The positions of the sharp minima coincide with the nuclear resonance conditions~\eqref{eq:NRC_even}.
		(b)~Revival amplitude $S^\perp_\mathrm{NESS}$ in the NESS (averaged over the last 10\% pulses) vs.\
		the magnetic field for various In concentrations $x$. The resonance conditions~\eqref{eq:NRC_even}~(dash-dotted lines) and~\eqref{eq:NRC_odd}~(dashed lines) for the various nuclear species in \InGaAs QDs are highlighted by vertical lines. The horizontal dashed line marks the SML steady state value as a reference to assess the effect of NIFF.}
	\label{fig:Sperp_Bext_np}
\end{figure}

The magnetic field dependence of the revival amplitude
\begin{align}
S^\perp(\np) \coloneqq \sqrt{[S^y(\np \TR^-)]^2 + [S^z(\np \TR^-)]^2} \,, \label{eq:Sperp}
\end{align}
vs.\ the number of applied pulses $\np$ is displayed in fig.~\ref{fig:Sperp_Bext_np}(a) for the In concentration $x=0.3$. The notation $\TR^-$ means that the spin polarization is probed just before the pump pulse.
The revival amplitude starts at $S^\perp = 0$ but rises quickly to the SML steady state value ${\SSML \approx 0.077}$ within $\mathcal{O}(10)$ pulses~\cite{beuge16,jasch17,klein18,scher20}. 
At this stage, the Overhauser field is still normally distributed. Subsequently, the revival amplitude changes 
due to the slowly emerging NIFF over a timescale covered by millions of pulses, eventually approaching a NESS.
The deviation of the amplitude from the SML steady state is representative for the degree of NIFF~\cite{scher20} in most cases, exceptions are discussed below.
The number of pulses $n_\mathrm{p,NESS}$ necessary to reach the NESS scales $\propto\Bext^2$~\cite{scher18,scher20}, e.\,g., for $\Bext = 10\,$T one must apply ${n_\mathrm{p,NESS} = 1.8\cdot 10^8}$ pulses.
Generally, the dependence of the revival amplitude on the external magnetic field is nonmonotonic.
The amplitude is large for small fields, then decreases with a broad minimum in the vicinity of $\Bext = 4.5\,$T, and increases thereafter. For fields larger than $\Bext\approx8\,$T, the amplitude decreases again. 
Most importantly, several sharp minima are discernible at larger fields.

The behavior becomes clearer when focusing on the NESS shown in fig.~\ref{fig:Sperp_Bext_np}(b) for various In concentrations $x$. The sharp minima are located exactly at the positions determined by the 
nuclear magnetic resonances~\eqref{eq:NRC_even} for $n=1$, i.\,e., 
whenever a nuclear Larmor period equals the pulse repetition time~$\TR$. 
The broad minimum is located in the range of magnetic fields allowing for half-integer nuclear resonances~\eqref{eq:NRC_odd} with $n=1$. 
It appears to be a robust feature as it is also observed in previous studies~\cite{klein18,scher20} 
where only a single nuclear spin species is considered, and a similar feature is also found experimentally for various QD samples~\cite{jasch17,klein18}.
The key difference to previous studies is the increased number of sharp minima and the much more complex behavior of the revival amplitude for large fields.
The depths of these minima strongly depend on the In concentration. Only the minimum stemming from the nuclear spin of As at around $10.4\,$T is almost independent of it as was to be expected because
varying $x$ does not alter the number of As isotopes in the QDs.
The more In is present (larger $x$), the deeper and sharper is the minimum related to the nuclear spin of
In at $\Bext = 8.072\,$T. In return, the two minima related to Ga become deeper when the In concentration is decreased.
Even for a small In concentration $x=0.0\overline{3}$, In prevails over the other isotopes due to its larger spin and larger hyperfine coupling strength, see table~\ref{tab:nuclei_parameters}.
Obviously, the sharp minimum at $\Bext = 8.072\,$T vanishes completely without any In ($x=0$). 
Likewise, the sharp minima related to the two Ga isotopes at $\Bext = 5.819\,$T and $7.403\,$T 
vanish without any Ga at $x=1$. In this case, the maximum of the revival amplitude around $8\,$T increases 
because the $n=2$ half-integer resonance~\eqref{eq:NRC_odd} related to $^{71}$Ga at around $8.7$\,T plays no role.
When Ga is present in the QD, this half-integer resonance limits the NIFF in this regime,
similarly to experimental observations for large fields~\cite{klein18}.
Generally, we find that the magnetic field dependence of the revival amplitude is fairly independent of the QD composition up to about $5\,$T. For larger fields, however, the behavior is complex and strongly depends on the In concentration~$x$. In this regime, various nuclear magnetic resonances influence the NESS and 
appear as sharp dips in the revival amplitude. These sharp resonance suggest themselves to be
exploited for a so far not known NMR spectroscopy.

\begin{figure}[t!]
	\centering
	\includegraphics[width=\columnwidth]{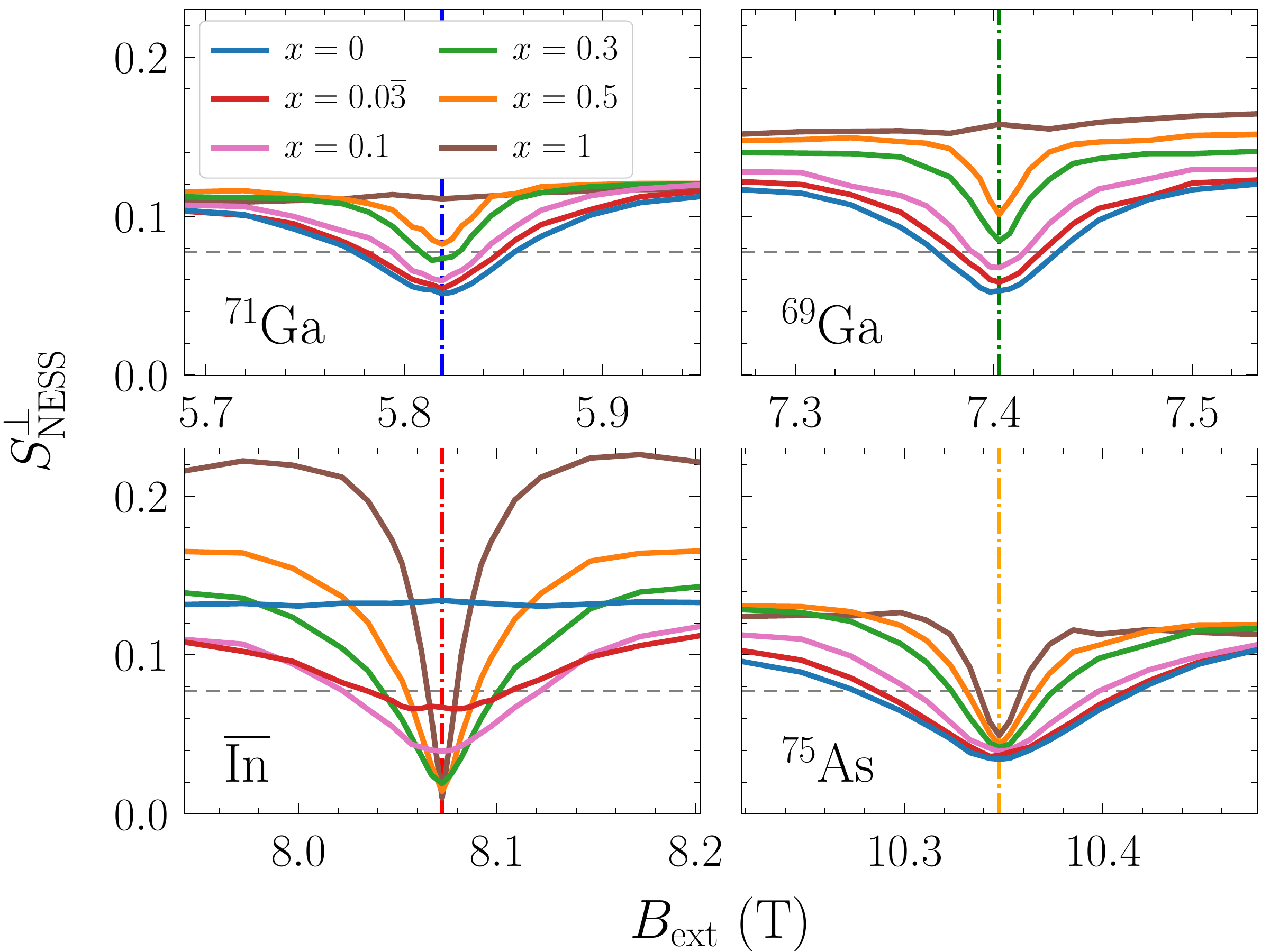}
	\caption{Revival amplitude $S^\perp_\mathrm{NESS}$ in the vicinity of the NMR fields
	fulfilling~\eqref{eq:NRC_even} for $n=1$ in the NESS for various In concentrations $x$. The panels show 
	$S^\perp_\mathrm{NESS}$ in the range $\pm 0.13\,$T around the NMR fields (dash-dotted lines) for $^{71}$Ga, $^{69}$Ga, $\overline{\mathrm{In}}$, and $^{75}$As. The SML steady state value is marked as horizontal dashed line.}
	\label{fig:Slim_NMR}
\end{figure}

\begin{figure}[t!]
	\centering
	\includegraphics[width=\columnwidth]{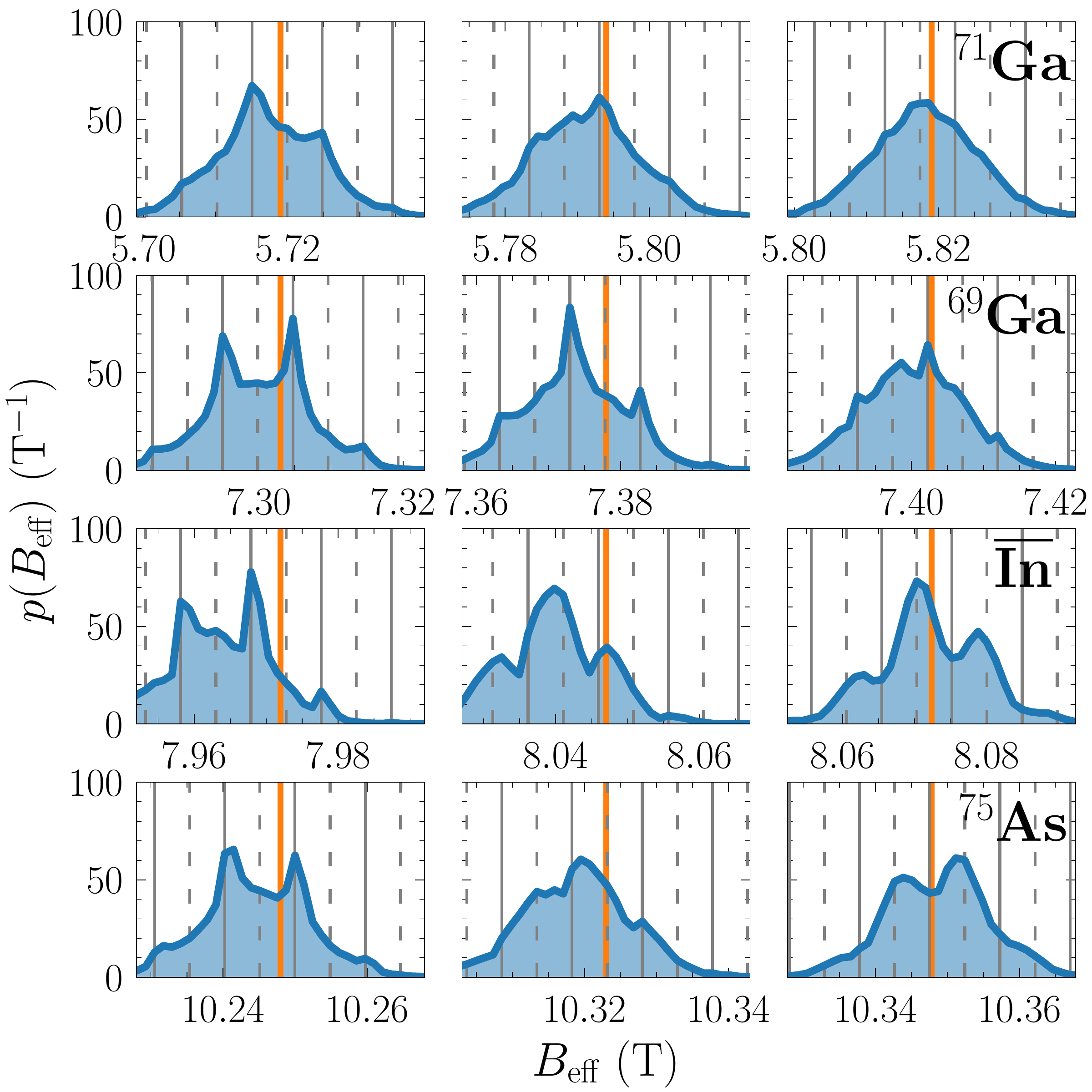}
	\caption{Probability distribution $p(\Beff)$ of the effective magnetic field in the NESS around the NMR fields
	fulfilling~\eqref{eq:NRC_even} for the In concentration $x=0.3$. Each row corresponds to the NMR for $^{71}$Ga, $^{69}$Ga, $\overline{\mathrm{In}}$, and $^{75}$As, respectively. In the right panels, $p(\Beff)$ is shown at the NMR field, in the center the applied field is lowered by $0.025$\,T, and on the right by $0.1\,$T. The applied field $\Bext$ is highlighted 
	in orange. The solid and dashed gray lines represent the phase synchronization conditions~\eqref{eq:ERC} 
	and~\eqref{eq:ORC} for the electron spin, respectively.}
	\label{fig:Overhauser_NMR}
\end{figure}

Figure~\ref{fig:Slim_NMR} depicts the revival amplitude in the NESS around the fields
meeting the nuclear resonance condition~\eqref{eq:NRC_even}. Obviously, the widths of the minima are related to the amount of In in the QD: a larger $x$ corresponds to a narrower minimum.
This is also the case for the dip related to As even though the number of As isotopes does not change with $x$.

The deviation of the revival amplitude from the SML value does not necessarily represent the degree of NIFF 
for fields close to the NMR. 
Here, the probability distribution of the effective field 
does not single out the resonances~\eqref{eq:ERC} or~\eqref{eq:ORC}; instead the peaks appear at irregular positions. 
This can be seen in fig.~\ref{fig:Overhauser_NMR} where the probability distribution $p(\Beff)$ is shown for external fields approaching the NMRs for the In concentration $x=0.3$.
For In and As, a change of resonance from~\eqref{eq:ERC} to~\eqref{eq:ORC} occurs. In contrast, the degree of NIFF at the resonance~\eqref{eq:ERC} 
for the two Ga isotopes decreases upon approaching the respective NMR field.
The details of the behavior depend on the In concentration, e.\,g., for smaller $x$ there is also a change of resonance visible for Ga.

Slight shifts of the mean value of the probability distributions $p(\Beff)$ from its initial value $\Bext$ 
are discernible in figs.~\ref{fig:SpinDynamics_Overhauser}(b) and~\ref{fig:Overhauser_NMR}. 
They stem from dynamic nuclear polarization, i.\,e., from the formation of a nonzero polarization of the nuclear spins parallel to the external field.
This effect is beyond the scope of this work, but has been observed in simulations before~\cite{scher20} and 
experimental hints at its existence can be found in ref.~\cite{jasch17}.

Finally, let us address whether and how the predicted features are experimentally accessible.
The two \InGaAs QD samples studied in ref.~\cite{klein18}, which were thermally annealed at different temperatures, have an estimated In concentration of $x \approx 0.3$~(sample~1) and $x \approx 0.4$~(sample~2) with most of the In in the center of the QDs~\cite{petrov08,sokolov16}. Thus, all nuclear species 
are present, but in these experiments the applied field was changed in steps of
$0.5\,$T which is too coarse to reveal sharp features. In principle, however, the external field can be tuned with 
mT accuracy so that sharp features such as those in fig.~\ref{fig:Slim_NMR} can be revealed. 
We stress that the inhomogeneity of a QD ensemble is not detrimental because the 
essential gyromagnetic ratios $\gamma_{\mathrm{n},k}$ do not depend on details of the QDs
and the electronic $g$ factor, which has a small spread in a real QD ensemble, does not influence the nuclear resonances.
Further experimental support can be obtained by applying radiation at the frequency specific for the particular kind of nuclear spins at a fixed magnetic field. Such related experiments have been performed recently~\cite{evers18} and it was found that this procedure also leads to a substantial reduction of the revival amplitude.

\section{Conclusions}
By methodical progress we were able to simulate the localized electron spins in quantum dots in presence
of a nuclear spin bath comprising several isotopes, hence several different nuclear spin species.
The key step was to separate fast and slow spin dynamics by an expansion in the
inverse of the applied magnetic field. Numerically, one only needs to integrate the slow dynamics
which runs two orders of magnitude faster than the brute force approach.
Clearly, this technique can be applied also to many other related physical problems. For instance, if there is only a single kind of nuclear spin the nonequilibrium dynamics between two pulses can be obtained algebraically. 
This enables extremely efficient simulations and can be used to deal with other issues such as 
inhomogeneously broadened trion transition energies
as they appear in real quantum dot~(QD) ensembles, adding another statistical component, or
with two-color pump-probe experiments~\cite{yugov09,glazov10,varwig14}. 

We studied the nonequilibrium dynamics of the localized electron spins in \InGaAs QDs subjected to extremely long trains of laser pulses inducing optical spin orientation and the controlled creation of nonequilibrium steady states~(NESSs).
These states are imprinted in the nuclear spin bath with a macroscopic lifetime~\cite{greil07a}.
Due to spin mode locking combined with nuclei-induced frequency focusing, a revival of the dephased spin polarization emerges before the arrival of the next pulse. Several resonances related to the Larmor precession of 
the different nuclear spins in the QDs determine the behavior decisively.
Especially, whenever the Larmor period of a nuclear spin matches the pulse repetition time the amplitude of the revival becomes minimal, giving rise to a number of very sharp minima in the magnetic field dependence. We suggest that these minima, stemming from nuclear magnetic resonances~(NMR), are observable in tailored experiments
constituting a novel kind of NMR spectroscopy.
Similar NESSs have been proposed to allow for the distillation of quantum coherent states~\cite{uhrig19}.

\acknowledgments
We thank P.~W.~Scherer for preliminary calculations, A.~Greilich for estimating the In concentration of the QD samples, and E.~Evers for discussing potential experimental realizations.
We gratefully acknowledge the resources provided by the Gauss Centre for Supercomputing e.V. on the supercomputer HAWK at High-Performance Computing Center Stuttgart
and by the TU Dortmund University on the HPC cluster LiDO3, partially funded by the German Research Foundation~(DFG) in project 271512359.
This work has been financed by the DFG and the Russian Foundation for Basic Research in the International Collaborative Research Centre TRR~160 in projects A4~and~A7.

\bibliographystyle{eplbib}

\end{document}


\makeatletter
\renewcommand{\fnum@figure}{Supplementary~\figurename~\thefigure}
\makeatother	
\renewcommand{\theequation}{S\arabic{equation}}

\maketitle

\section{Fast and slow degrees of freedom}
\label{app:O1h}

The methodical key achievement is the separation of fast and slow degrees of freedom.
It amounts up to an expansion in the inverse external magnetic field.
Only this progress has made the thorough study of the nonequilibrium spin physics in QDs 
consisting of different isotopes possible.
The fast degrees of freedom are treated analytically while the slow degrees are treated 
in a controlled expansion in~$\Bext^{-1}$ and generically solved numerically
by integration of the system of differential equations.
The equations of motion~(EoMs) of the main text
\begin{subequations}
	\begin{align}
	\ddt \S &= ( \Bov + \gamma_\mathrm{e} B_\mathrm{ext} \ex ) \times \S + \frac{1}{\tau_0} J^z \ez \,,\\
	\ddt \bm{B}_{\mathrm{Ov},k} &= ( A_k \S + \gamma_{\mathrm{n},k} B_\mathrm{ext} \ex) \times \bm{B}_{\mathrm{Ov},k} \,,
	\end{align}
	\label{eq:EoM_isotopes}%
\end{subequations}
are split into a part describing the fast Larmor precession 
and another part describing the slow dynamics by applying suitable ans\"atze. 
Corrections of the order $\mathcal{O}(\Bext^{-2})$ are neglected because they are strongly suppressed.

To this end, we switch to the composite variables
\begin{subequations}
	\begin{align}
	z &\coloneqq S^y + \mathrm{i} S^z\,,\\
	b_k &\coloneqq B^y_{\mathrm{Ov},k} + \mathrm{i} B^z_{\mathrm{Ov},k}\,,\\
	B^x_k &\coloneqq B^x_{\mathrm{Ov},k}\,,
	\end{align}
\end{subequations}
and use the abbreviations $h = \gamma_\mathrm{e} \Bext$, $\hnk = \gamma_{\mathrm{n},k} \Bext$, and $\hh = \gamma_\mathrm{h} \Bext$ for the Larmor frequencies. In this notation, the EoMs~\eqref{eq:EoM_isotopes} have the form
\begin{subequations}
	\begin{align}
	\ddt z &= \mathrm{i} (B^x + h)z - \mathrm{i} b S^x + \mathrm{i} \frac{1}{\tau_0} J^z(t)\,, \\
	\ddt b_k &= \mathrm{i} (A_k S^x + \hnk) b_k - \mathrm{i} A_k z B_k^x\,, \\
	\ddt S^x &= \Im(z b^\ast)\,,\\
	\ddt B_k^x &= - A_k \Im(z b_k^\ast)\,,
	\end{align}
	\label{eq:EoM_isotopes_2}%
\end{subequations}
with $b = \sum_k b_k$, $B^x = \sum_k B^x_k$, and
\begin{align}
	J^z(t) = J^z(0)\, \mathrm{e}^{-t/\tau_0} \cos(\hh t)\,.
\end{align}

\subsection{Equations of motion in $\mathcal{O}(\Bext^{-1})$}

The ans\"atze
\begin{subequations}
	\begin{align}
	z(t) &= z_0(t) + z_1(t) \eiht + z_2(t) \eihht + z_3(t) \emihht \,,\\
	S^x(t) &= S^x_0 (t) + \Re\left[S^x_1(t) \eiht \right] \nonumber \\
	&= S^x_0 (t) + \frac{1}{2} \left( S^x_1 \eiht + S^{x\ast}_1 \emiht \right)\,,\\
	b_k(t) &= b_{0,k}(t) + b_{1,k}(t) \eiht \,,\\
	B^x_k(t) &= B^x_{0,k} (t) + \Re\left[B^x_{1,k}(t) \eiht \right] \nonumber \\
	&= B^x_{0,k} (t) + \frac{1}{2} \left( B^x_{1,k} \eiht + B^{x\ast}_{1,k} \emiht \right)\,,
	\end{align}
	\label{eq:ansatze}%
\end{subequations}
which already include the relevant terms stemming from the fast Larmor precession with frequencies $h$ and $\hh$, are inserted into eqs.~\eqref{eq:EoM_isotopes_2}.
We identify and keep all terms which are $\mathcal{O}(1)$ or $\mathcal{O}(h^{-1})$ and impose
 $\mathcal{O}(\hh) = \mathcal{O}(h)$ and $\mathcal{O}(\hn) = \mathcal{O}(1)$.
The first relation is justified because the $g$ factors of electrons and holes are
of similar magnitude ($g_\mathrm{e} = 0.555$~\cite{greil07a}, $g_\mathrm{h} = 0.15$~\cite{yugov07}) while the second relation is justified because the gyromagnetic ratios of the nuclei in \InGaAs QDs are smaller by three orders of magnitude than the electronic ones~\cite{coish09}.
In $\Oih$ the resulting relations are algebraic,
\begin{subequations}
	\begin{align}
	z_0(t) &= \frac{1}{h} b_0 S^x_0 \,, \\
	z_2(t) &= -\frac{J^z(0)}{2 \tau_0 (h - \hh)} \,\mathrm{e}^{-t/\tau_0} \,,\\
	z_3(t) &= -\frac{J^z(0)}{2 \tau_0 (h + \hh)} \,\mathrm{e}^{-t/\tau_0} \,,\\
	b_{1,k}(t) &= - \frac{A_k}{h} z_1 B^x_{0,k} \,, \\
	S^x_1(t) &= - \frac{1}{h} z_1 b_0^\ast \,, \\
	B^x_{1,k}(t) &= \frac{A_k}{h} z_1 b_{0,k}^\ast \,, \\
	S^x_0(t) &= S^x_0(0) \,,
	\end{align}
	\label{eq:algebraic_relations_isotopes}%
\end{subequations}
while in $\mathcal{O}(1)$ we obtain the differential equations
\begin{subequations}
	\begin{align}
	\ddt z_1 &= \mathrm{i} \left( B^x_0 z_1 - b_1 S^x_0 + \frac{1}{2} b_0 S^x_1 \right) \,, \\ 
	\ddt b_{0,k} &= \mathrm{i} \left[ (A_k S^x_0 + \hnk) b_{0,k} - A_k z_0 B^x_{0,k} - \frac{A_k}{2} z_1 B^{x\ast}_{1,k} \right] \,, \\
	\ddt B^x_{0,k} &= - A_k \Im ( z_0 b_{0,k}^\ast) \,.
	\end{align}
	\label{eq:DEQ_isotopes}%
\end{subequations}
It turns out that $z_1$, $b_{0,k}$, $S^x_0$, and $B^x_{0,k}$ are of order $\mathcal{O}(1)$ while the remaining variables 
$z_0$, $z_2$, $z_3$, $b_{1,k}$, $S^x_1$, and $B^x_{1,k}$ are of order $\mathcal{O}(h^{-1})$. All other conceivable corrections are of order $\mathcal{O}(h^{-2})$ and thus strongly suppressed for the large magnetic fields studied in the main text.
Hence, they can be omitted.

By inserting the algebraic relations~\eqref{eq:algebraic_relations_isotopes} into eqs.~\eqref{eq:DEQ_isotopes}, we obtain
\begin{subequations}
	\begin{align}
	\ddt z_1 &= \mathrm{i} \left( B^x_0 + \sum_k \frac{A_k}{h} B^x_{0,k} S^x_0 + \frac{|b_0|^2}{2h} \right) z_1 \,, \label{eq:ddt_z1_isotopes}\\
	\ddt b_{0,k} &= \mathrm{i} \left( A_k S^x_0 + \hnk - \frac{A_k^2}{2h} |z_1|^2 \right) b_{0,k} \nonumber\\ &\quad - \mathrm{i} \frac{A_k}{h} S^x_0 B^x_{0,k} b_0 \,, \label{eq:ddt_b0k_isotopes}\\
	\ddt B^x_{0,k} &= - \frac{A_k}{h} S^x_0 \Im \left( \sum_{j \ne k} b_{0,j} b_{0,k}^\ast \right) \,. \label{eq:ddt_Bx0k_isotopes} 
	\end{align}	
\end{subequations}
Equation~\eqref{eq:ddt_z1_isotopes} can be simplified by applying the complex exponential ansatz
\begin{align}
	z_1(t) = z_1(0) \,\mathrm{e}^{\mathrm{i} \varphi_{z_1}(t)} \,,
\end{align}
which yields the new differential equation
\begin{align}
\ddt \varphi_{z_1} = B^x_0 + \sum_k \frac{A_k}{h} B^x_{0,k} S^x_0 + \frac{|b_0|^2}{2h} \,. \label{eq:ddt_phiz1_isotopes}
\end{align}
The set of differential equations~\eqref{eq:ddt_b0k_isotopes}, \eqref{eq:ddt_Bx0k_isotopes}, and~\eqref{eq:ddt_phiz1_isotopes} are the ones to be solved numerically starting from appropriate initial conditions.

Remarkably, if only a single nuclear species is present the sums in the equations vanish such that the differential equations can be easily solved analytically. The dynamics for this simpler case is therefore fully determined by a set of algebraic relations in order $\mathcal{O}(h^{-1})$.

\subsection{Initial conditions}
We insert $t = 0$ into the algebraic relations~\eqref{eq:algebraic_relations_isotopes} to determine the initial conditions of all time-dependent variables.
For brevity, the time argument $t=0$ is omitted in the following.
The initial conditions for the quantities $z$, $b_k$, $S^x$, $B_k^x$, and $J^z$ are known: $z$,~$S^x$, and $J^z$ are determined by the nondeterministic pulse whereas $b_k$ and $B_k^x$ representing the Overhauser subfields are determined by their Gaussian initial conditions.
The task is to deduce from them the initial conditions of the expansion variables 
$z_0$, $z_1$, $b_{0,k}$, $b_{1,k}$, $S^x_0$, $S^x_1$, $B^x_{0,k}$, and $B^x_{1,k}$.
First, we define the auxiliary quantities
\begin{align}
P_k \coloneqq \Re ( z_1 b_{0,k}^\ast )
\end{align}
in which we insert the relation
\begin{align}
b_{0,k} = b_k - b_{1,k} = b_k + \frac{A_k}{h} z_1 B^x_{0,k} \,. 
\end{align}
After a few algebraic transformations, we obtain
\begin{align}
P_k = \frac{\Re(z_1 b_k^\ast) + \frac{A_k}{h} B^x_k |z_1|^2}{1 + \frac{A_k^2}{h^2} |z_1|^2} \,,
\end{align}
which depends only on known initial conditions except the initial value $z_1$.
Furthermore, we have
\begin{subequations}
	\begin{align}
	B^x_{0,k} &= B^x_k - \frac{A_k}{h} P_k \,,\\
	S^x_0 &= S^x + \frac{1}{h} \sum_k P_k \,,
	\end{align}
\end{subequations}
which we insert into $z_1 = z - z_0 - z_2 - z_3$ and find
\begin{align}
z_1 \nonumber &= z \\ &- \frac{1}{h} \sum_k \left[ b^k - \frac{A_k}{h} z_1 
\left( B^x_k - \frac{A_k}{h} P_k \right) \right] \left(S^x + \frac{1}{h} \sum_j P_j \right) \nonumber 
\\ &+ \frac{J^z}{2\tau_0} \left(\frac{1}{h+\hh} + \frac{1}{h-\hh} \right) \,.
\end{align}
Since $P_k$ depends on $z_1$, this nonlinear and nonpolynomial relation determines $z_1$. 
All terms on the right-hand side except $z$ are suppressed by at least a factor $h^{-1}$.
Hence, the relevant zero for $z_1$ is quickly found by iteration starting from $z_1 = z$.
Convergence is reached within $\mathcal{O}(5)$ iteration steps.
Then, all initial conditions are fully determined; the remaining ones are
\begin{subequations}
	\begin{align}
	z_0 &= z - z_1 - z_2 - z_3 \,,\\
	z_2 &= -\frac{J^z}{2 \tau_0 (h - \hh)} \,,\\
	z_3 &= -\frac{J^z}{2 \tau_0 (h + \hh)} \,, \\
	b_{1,k} &= -\frac{A_k}{h} z_1 B^x_{0,k} \,,\\
	S^x_1 &= - \frac{1}{h} z_1 \sum_k b_{0,k}^\ast \,, \\
	B^x_{1,k} &= \frac{A_k}{h} z_1 b_{0,k}^\ast \,.
	\end{align}
\end{subequations}

\subsection{Validity of the expansion}

\begin{figure}[t!]
	\centering
	\includegraphics[width=\columnwidth]{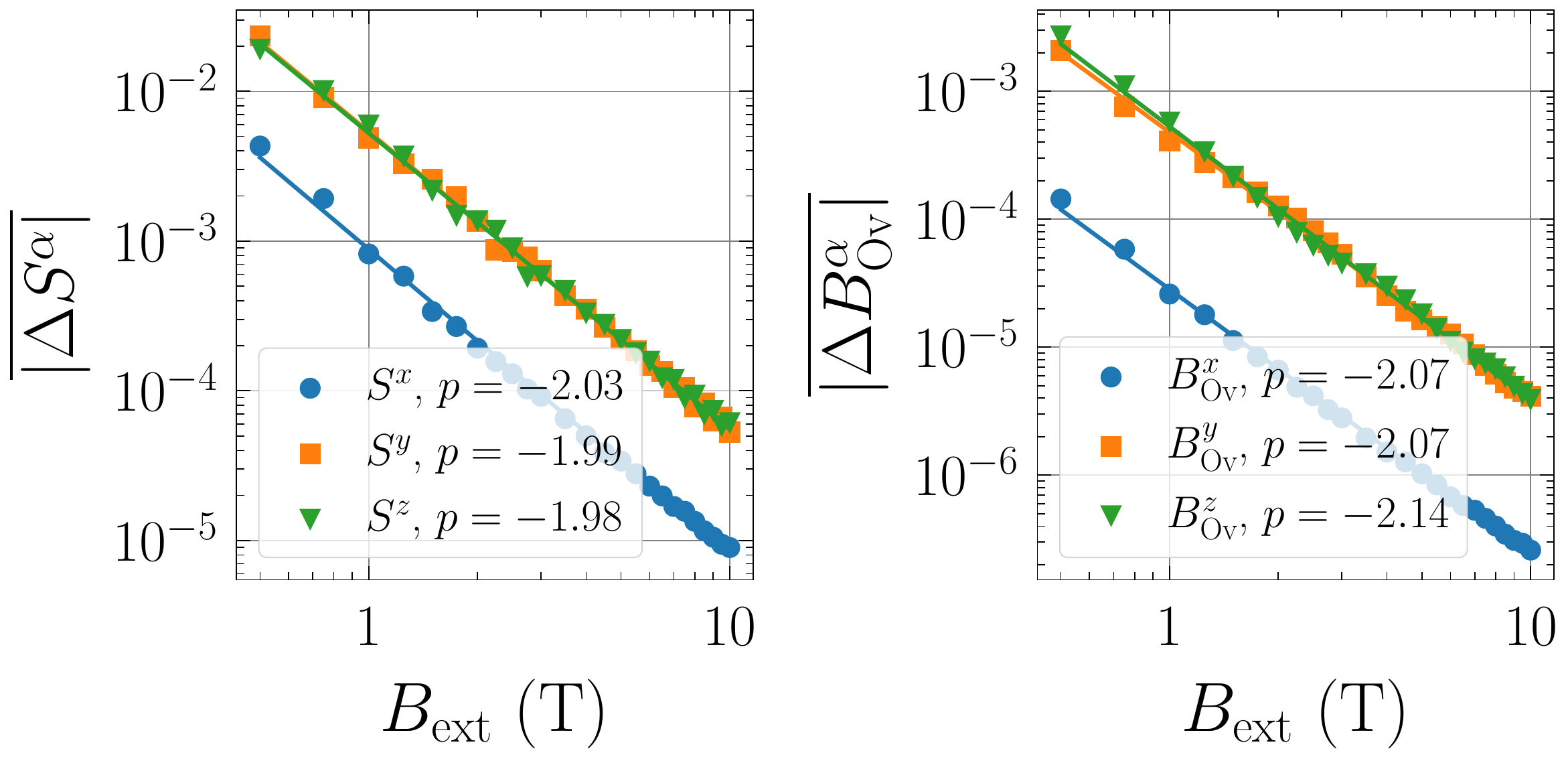}
	\caption{Error analysis of the $\mathcal{O}(\Bext^{-1})$ expansion for the electron spin $\S$ and the Overhauser field $\Bov$. The errors averaged over $M=256$ random initial conditions (denoted by the overline) at $t = \TR$, \textit{i.e.}, after the first pulse period, is shown in a double logarithmic plot. The slopes $p$ of the applied linear fits (solid lines) are given in the legends; they all have a fit error of $\pm\,0.02$. Clearly, the error is of $\mathcal{O}(\Bext^{-2})$ for all components. Parameters: $N=200$, $\Tnstar = 1\,$ns, $x = 0.3$.}
	\label{fig:isotopes_error}
\end{figure}

We calculate the time evolution of $\S$ and $\Bov$ for $M=256$ random initial conditions by solving the original EoM~\eqref{eq:EoM_isotopes} numerically and also applying the $\mathcal{O}(\Bext^{-1})$ expansion. 
The average absolute error calculated for each component at $t = \TR = 13.2\,$ns is shown in suppl.~fig.~\ref{fig:isotopes_error} in a double logarithmic plot. Linear fits reveal a clear $\Bext^{-2}$ dependence for all errors.
This confirms the accuracy of the derived equations to be of order~$\mathcal{O}(\Bext^{-1})$.
	
\begin{figure}[t!]
	\centering
	\includegraphics[width=\columnwidth]{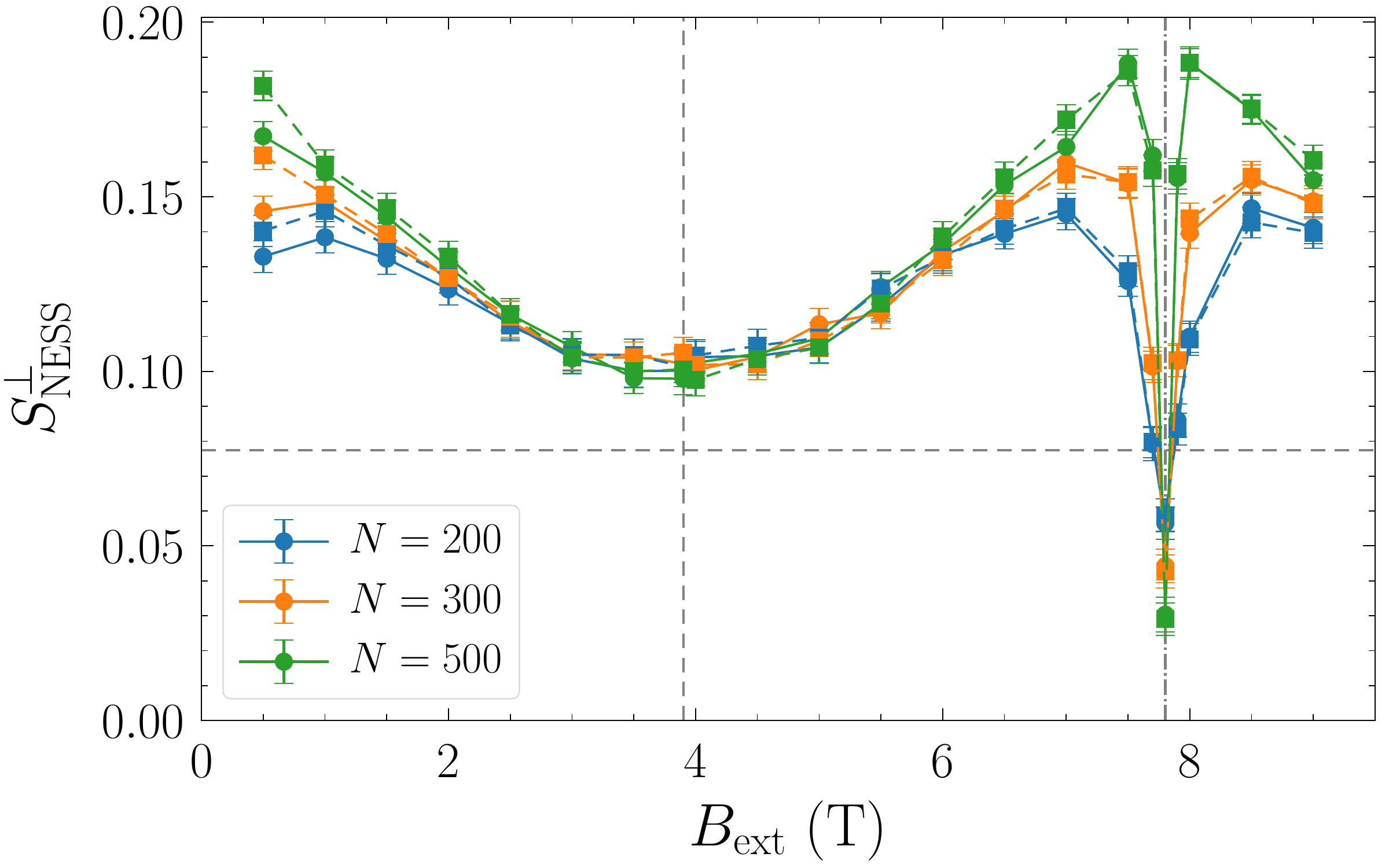}
	\caption{Revival amplitude $S^\perp_\mathrm{NESS}$ as a function of the magnetic field $\Bext$ in the nonequilibrium steady state~(NESS) regime for the model where only a single nuclear isotope is considered. The numerically exact
	(spheres, solid lines) and the approximate $\mathcal{O}(\Bext^{-1})$ (squares, dashed lines) results
	are shown for various numbers of nuclear spins $N$. The agreement is within the statistical accuracy for magnetic fields $\Bext \gtrsim 1\,$T. 
	Parameters: $\Tnstar = 1\,$ns, $\gamma_{\mathrm{e}}/\gamma_{\mathrm{n}} = 800$, $I=3/2$, $M=10240$.}
	\label{fig:Slim_Bext_single_isotope_comparison}
\end{figure}

Supplementary~Figure~\ref{fig:Slim_Bext_single_isotope_comparison} shows the accuracy of the approach for the magnetic field dependence of the revival amplitude in the nonequilibrium steady state~(NESS) for the simple model where only a single isotope ($\gamma_{\mathrm{e}}/\gamma_{\mathrm{n}} = 800$) is considered. For magnetic fields $\Bext \gtrsim 1\,$T, the accuracy of the $\mathcal{O}(\Bext^{-1})$ approach is within the statistical accuracy. The revival amplitude in the NESS regime is calculated as the arithmetic mean over the last $10\%$ pulses, the displayed error is the corresponding standard deviation.

\begin{figure}[b!]
	\centering
	\includegraphics[width=\columnwidth]{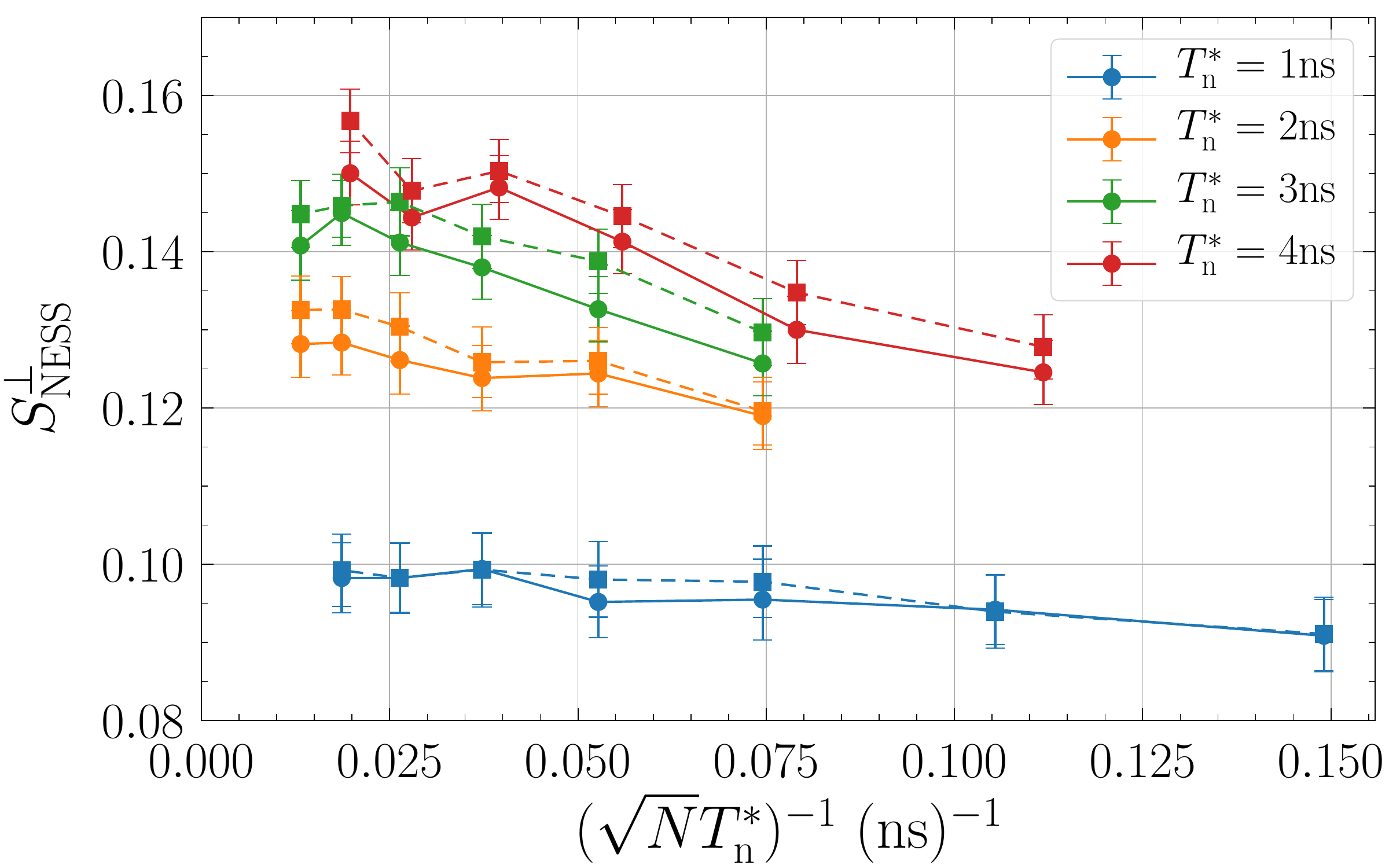}
	\caption{Revival amplitude $S^\perp_\mathrm{NESS}$ as a function of the effective coupling strength $(\sqrt{N}\Tnstar)^{-1}$ at a magnetic field of $\Bext = 1\,$T. The numerically exact (spheres, solid lines) and approximate $\mathcal{O}(\Bext^{-1})$ (squares, dashed lines) results are shown for various dephasing times $\Tnstar$. The agreement is within the statistical accuracy, but the approximate solution slightly overestimates the real revival amplitude. Parameters:~$x = 0.3$, $M=11520$.}
	\label{fig:Slim_Bext_N_Tnstar}
\end{figure}

We also compare the numerically exact and approximate $\mathcal{O}(\Bext^{-1})$~results for the revival amplitude as a function of the number of nuclei $N$ for several dephasing times $\Tnstar$ in suppl.~fig.~\ref{fig:Slim_Bext_N_Tnstar} for an In$_{0.3}$Ga$_{0.7}$As QD sample. Due to computational constraints for the exact numerics, the comparison is only carried out for $\Bext = 1\,$T. The agreement between the numerically exact and approximate results is within the statistical accuracy for all combinations of $N$ and $\Tnstar$, but the approximate solution slightly overestimates the real revival amplitude for this rather small magnetic field. For larger fields, the agreement is expected to be better due
to the construction based on a systematic expansion in $\Bext^{-1}$; see also suppl.~fig.~\ref{fig:Slim_Bext_single_isotope_comparison} where the accuracy improves for larger fields.

Physically, the results shown in suppl.~fig.~\ref{fig:Slim_Bext_N_Tnstar} demonstrate the influence of the dephasing time on the revival amplitude for the model studied in the main text. 
For $\Tnstar = 1\,$ns, the revival amplitude is only slightly larger than the SML steady state value ${\SSML = \nicefrac{1}{\sqrt{3}} - \nicefrac{1}{2} \approx 0.077}$, \textit{i.e.}, the degree of NIFF is small.
For the value $\Tnstar = 4\,$ns studied in the main text, there is a noticeable dependence on the number of nuclei. 
For $N=60$ as used in the main text, we find that the revival amplitude is already close to its maximum value, \textit{i.e.}, 
it will not change significantly for larger values of~$N$. 
For the case of a single nuclear species treated in ref.~\cite{scher20}, the choice of $\Tnstar$ had 
no significant effect in the limit of an infinite bath size $N\to\infty$, which is the experimentally relevant situation due to the $N=10^4 - 10^6$~\cite{urba13} effectively coupled nuclear spins in a QD.
In contrast, for the case of several nuclear species
treated in this study there is a significant influence of $\Tnstar$
on the revival amplitude, even in the limit of an infinite bath size $N\to\infty$.
We emphasize that the value $\Tnstar=4\,$ns is chosen in accordance with previous analyses~\cite{greil06a,fischer18}.

\bibliographystyle{eplbib}